\documentclass[11pt,dvips]{article}
\usepackage{epsfig,times} 
\usepackage{picinpar}
\usepackage{wrapfig}
\usepackage{floatflt}
\makeatletter
\renewcommand{\section}{\@startsection{section}{1}{0in}
	{0.4\baselineskip}{0.1\baselineskip}{\Large\bf}}
\renewcommand{\subsection}{\@startsection{subsection}{2}{0in}
	{0.25\baselineskip}{-\baselineskip}{\large\bf}}
\renewcommand{\subsubsection}{\@startsection{subsubsection}{3}{0in}
	{0.1\baselineskip}{-\baselineskip}{\normalsize\bf}}
\makeatother
\def\etal{{\it et al.~}}

\def\eg{{\it e.g.,~}}

\def\ltsima{$\; \buildrel < \over \sim \;$}
\def\kmsmpc{~{$\rm km~s^{-1}~Mpc^{-1}$}}
\begin{document}

%
\thispagestyle{myheadings}
%
\markright{OG.3.3.31}
\begin{center}
%
{\LARGE \bf Large Cosmic Shock Waves as Sites for Particle Acceleration}
\end{center}

\begin{center}
%
%
{\bf Francesco Miniati$^{1}$, Dongsu Ryu$^{2}$, Hyesung Kang$^{3}$, 
and T. W. Jones$^{1}$}\\
{\it $^{1}$School of Physics and Astronomy, University of Minnesota,
    Minneapolis, MN 55455 \\
$^{2}$Department of Astronomy \& Space Science, Chungnam National
    University, Korea \\
$^{3}$ Department of Astronomy, Pusan National University,
Pusan, 609-735 Korea}
\end{center}

\begin{center}
{\large \bf Abstract\\}
\end{center}
\vspace{-0.5ex}
%
%
The properties of cosmic shock waves are studied through numerical 
simulations in two cosmological scenarios (SCDM and $\Lambda$CDM).  
The scaling relations for the average radius and velocity associated with the
accretion shocks are somewhat different,
yet qualitatively similar to the self similar solutions for a flat $\Omega_M=1$
universe. The energy supplied by infalling gas at accretion shock waves is 
large enough to sustain production of 
abundant cosmic ray populations if a viable acceleration mechanism can take
place there. Finally, in addition to shocks created by the encounter of the
merging ICMs of two clusters of galaxies, accretion shocks
associated with the merging clusters generate strong ``relic'' shocks which
propagate through the ICM producing additional heating of the ICM,
and associated CR acceleration.
\vspace{1ex}

%
%
\section{Introduction:}
\label{intro.sec}

Cosmic Rays (CRs) acceleration and transport in Galaxy Clusters (GCs) 
environments is an important topic for several reasons.
In fact, being characterized by a huge size ($\sim$ several Mpc) and 
very large Mach numbers ($\sim 10-10^3$, 
Miniati \etal 1999),
accretion cosmic shock waves have been proposed as sites for acceleration of
high energy CRs up to $10^{18} - 10^{19}$eV. Such energies
would be in fact achievable through 
Jokipii diffusion of the relativistic particles if about
$10^{-4}$ of the kinetic energy associated with the accretion flow can be
injected into CRs (Kang, Ryu \& Jones 1996;
Kang, Rachen \& Biermann 1997).

In addition, the recent detection of
diffuse excess EUV emission compared to that expected from the X-ray emitting 
Intra Cluster Medium (ICM) for all GCs observed by the EUVE satellite 
(Lieu \etal 1999, Bowyer, Lieu \& Mittaz 1998 and references therein)
suggested nonthermal
contributions much higher than previously expected.
The interpretation of the EUV excess as Inverse Compton (IC)
emission due to low energy relativistic electrons scattering off the cosmic
microwave background looks preferable to a thermal model 
for several reasons (Sarazin \& Lieu 1998).
However, while viable, a correct IC-based model which accounts for the
radio emission and the magnetic field data is not straightforward
(Sarazin \& Lieu 1998; Bowyer \& Bergh\"ofer 1998, Ensslin, Lieu \& Biermann
1999). From published studies a number of consequences emerge. 
Most dramatic, the electron
component of CRs could amount to a few percent of the
thermal energy content in GCs (Sarazin \& Lieu 1998). In turn, this implies
that proton CRs, due to their much longer cooling time, could 
contribute a large fraction of the total pressure there
(Lieu \etal 1999). This result could have a great
impact on cosmology, because GCs' mass estimates are usually based on the
assumption of hydrostatic equilibrium for the ICM gas in
the potential well of the total cluster mass. 
In general, it is clear that a single population 
of CRs cannot account for all the
observed nonthermal emission and that the relative distribution of CRs and
magnetic fields plays a crucial role in a correct interpretation
(Ensslin, Lieu \& Biermann 1999). 

Finally, Ryu, Kang \&
Biermann (1997) addressed the issue of topological characteristics of cosmic
magnetic
fields, concluding that since the magnetic flux is strongly aligned along 
cosmic structures such as ``sheets'' and ``filaments'', values \ltsima $1\mu$G
are permitted by current rotational measures (Kronberg 1994). In a follow-up
paper the same authors discussed the implications
of their results for the transport of Ultra High Energy Cosmic Rays
(Biermann, Kang \& Ryu 1996).

In this paper we describe the properties of shock
waves occurring as a result of accretion flows onto GCs and of cluster mergers
with respect to their potential contribution to CRs production. 

\section{Method}
\label{format.sec}

We have studied
the properties of shocks formed in numerical simulations of
large scale structure in the universe using the hydrodynamic 
code recently developed by Ryu \etal (1993). 
The simulations involved a cubic region of current epoch
size 85 $h^{-1}$ Mpc,
inside a computational box with $270^3$ cells and $135^3$ dark matter 
particles. This provides a spatial resolution of a few times 
0.31 $h^{-1}$ Mpc.

We investigate two particular cases: 1)
a standard Cold Dark Matter (SCDM) model (Kang \etal 1994) with
$h \equiv $ H$_0/(100$ \kmsmpc$) = 0.5$, $\Omega_0 = \Omega_M = 1$, 
baryonic fraction $\Omega_b = 0.06$, and 2) a
CDM + $\Lambda$ model (Cen \& Ostriker 1994) with 
$h = 0.6,~ \Omega_0 =\Omega_M +\Omega_\Lambda = 1, 
\Omega_M = 0.45, ~\Omega_\Lambda = 0.55, ~\Omega_b = 0.043$ (see Kang \etal
1994 and Cen \& Ostriker 1994 for more details). 
Finally, we refer to Miniati \etal (1999) for 
a description of the method used to select clusters and
to determine their temperature and accretion shock properties.

\section{Results}
\begin{wrapfigure}[14]{r}{8.8cm}
\vspace{4.2cm}
\includegraphics{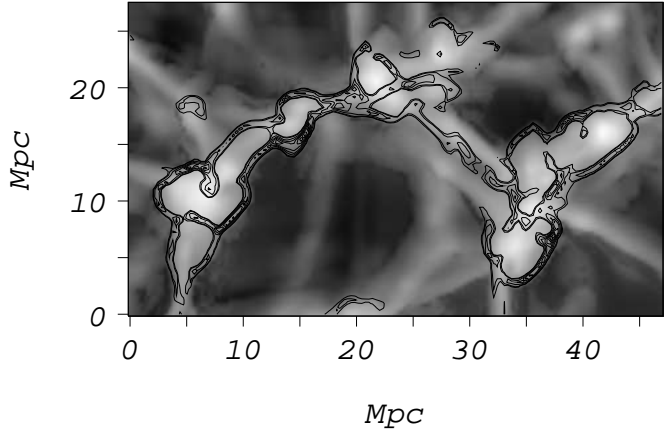}
\caption{\it A typical Cosmic Structure: shock waves contours superposed on
grayscale image of X-ray bremsstrahlung emission (text for details).}
\label{fig1}
\end{wrapfigure}
Fig. \ref{fig1} illustrates a slice of a typical cosmic structure.
It shows 
contours of compression ($\nabla \cdot v$)
corresponding to shock waves, superposed on a grayscale image of X-ray
bremsstrahlung emission from the hot ICM (brighter regions correspond to higher
emission). The figure shows shock waves forming around clusters and
filaments as a result of supersonic accretion flow. 
Shocks around clusters are responsible for the heating of the ICM and
several models have been proposed to describe their properties (Bertschinger
1984, Ryu \& Kang 1997). 
In particular, in self similar solutions for a $\Omega_M =1$ universe,
the following relations
between $v_a$ and $T_x$ as well as $r_a$ and $T_x$ are expected for accretion
shocks
(Bertschinger 1985):
\begin{eqnarray}
v_a &=& K_{v_a} ~\left(\frac{T_x}{7 \times 10^7\mbox{K}}\right)^{\alpha_{v_a}}
= 1.31 \times 10^3 \mbox{Km s$^{-1}$}
\left(\frac{T_x}{7 \times 10^7\mbox{K}}\right)^\frac{1}{2} \\
r_a &=& K_{r_a} ~\left(\frac{T_x}{7 \times 10^7\mbox{K}}\right)^{\alpha_{r_a}}
= 2.12~ h^{-1} \mbox{Mpc} 
\left(\frac{T_x}{7 \times 10^7\mbox{K}}\right)^\frac{1}{2}
\end{eqnarray}
In addition to stationary shocks around GCs,
shocks traveling through the ICM are also of great interest for their potential
contribution to heating the ICM and/or CR acceleration. 
Such shocks are commonly produced during a merger event at the
interface between two cluster. However, during the
merging process the accretion shocks associated with the initial clusters
also collide, producing strong ``relic'' shocks that propagate in the outer 
regions of the ICM of the new-born cluster. Over the evolutionary 
history of a cluster, several merging processes occur, creating
an ICM rich in shock structures (Miniati \etal 1999). 
Both merging and ``relic'' shocks are clearly visible in Fig. \ref{fig1} .

\begin{wrapfigure}[23]{r}{9.3cm}
\vspace{8.5cm}
\includegraphics{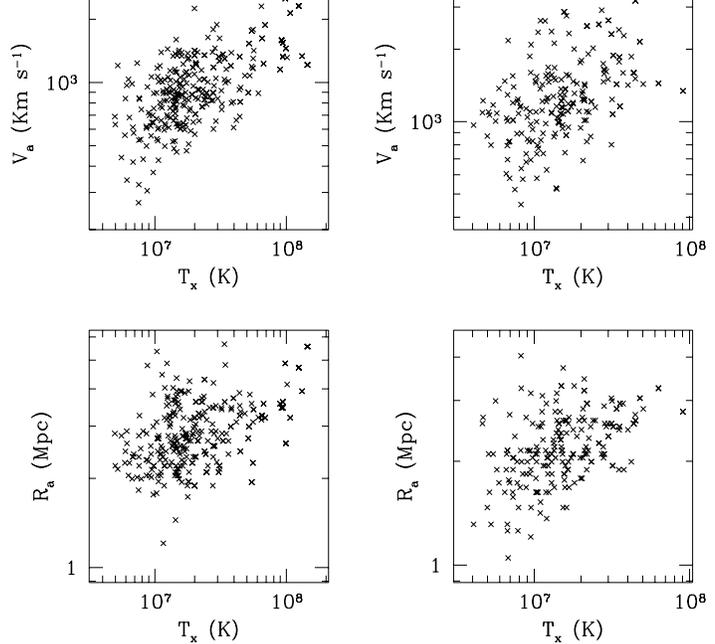}
\caption{\it Properties of accretion shock waves: average velocity (top) and
radius (bottom) as a function of average cluster core temperature for SCDM
(left) and $\Lambda$CDM (right).}
\label{fig2}
\end{wrapfigure}

Some quantitative properties of cosmic shocks,
as derived from our numerical study, are summarized in 
Fig. \ref{fig2}.
It shows the average accretion shock velocity (top panels) and radius (bottom
panels) as a function of the cluster core temperature (one of GCs best 
measured quantities), defined as the average temperature inside a central 
region of radius 0.5 $h^{-1}$ Mpc. 
The trend is qualitatively similar, but quantitatively different from the 
analytical 
relations of eq. 1 and 2. In the SCDM model 
we find best least square fit with 
$K_{v_a} = 1400$ Km s$^{-1}$, $\alpha_{v_a} = $ 0.33 and 
$K_{r_a} = 3.4$ Mpc, $\alpha_{r_a} = $ 0.15. 
While numerical errors both in the cosmological simulation and in the data
analysis must be considered, such strong deviations from the self
similar model should be not too surprising given the highly idealized
assumptions for cluster formation under which eq. 1 and 2 were derived.
The larger discrepancy for the accretion radius probably
reflects the complex shock structure in the simulation.
We mention for completeness that in the $\Lambda$CDM case we obtain
$K_{v_a} = 1960$ Km s$^{-1}$, $\alpha_{v_a} = $ 0.29 and 
$K_{r_a} = 2.8$ Mpc, $\alpha_{r_a} = $ 0.18. 
We note that for both the SCDM and $\Lambda$CDM, 
over the range of temperatures of the 
selected clusters ($T_x \simeq 6\times 10^6$K
- $10^8$K), the accretion velocity extends over at least one order of magnitude 
from a few $10^2$ Km s$^{-1}$ to a few $10^3$ Km s$^{-1}$. The shock radius, 
on the other hand, only varies by a factor of a few. Whereas the shock
speeds are comparable to those characterizing young supernova remnants (SNRs), 
shock sizes
associated with GCs are enormously greater. 
In addition, since the accretion flows around GCs are much colder than the
interstellar medium, Mach numbers associated with the
accretion shocks are much larger than for SNRs. In this respect we note further
that accretion shocks occurring in a $\Lambda$CDM scenario
have consistently much larger Mach numbers than in the SCDM (Miniati \etal 1999).
 
\begin{wrapfigure}[13]{r}{9.3cm}
\vspace{4.0cm}
\includegraphics{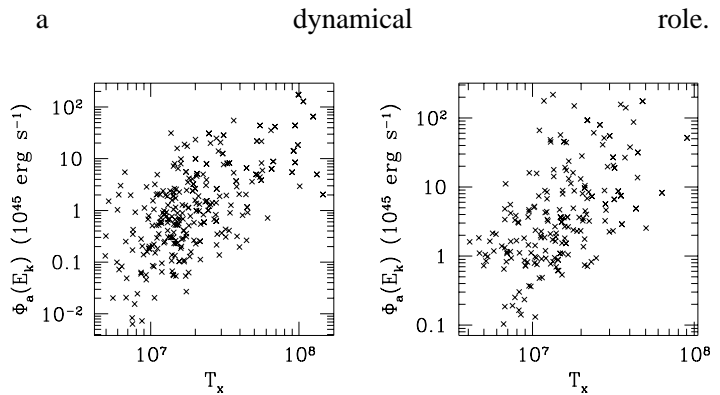}
\caption{\it Average kinetic energy flux at accretion shocks as a function of
cluster core temperature for SCDM
(left) and $\Lambda$CDM (right).}
\label{fig3}
\end{wrapfigure}

Another quantity of interest is the flux
of kinetic energy associated with the gas crossing the shock. This is given by
\begin{equation}
\Phi(E_k) = \frac{1}{2} ~\rho ~v_a^2 ~r_a^2 ~v_a ~\vert_{r=r_a},
\end{equation}
and is plotted, as a function of cluster core temperature, in Fig. \ref{fig3}. 
The flux 
$\Phi(E_k)$ supplies a large amount of energy, not surprisingly, 
comparable to the cluster X-ray luminosity. 
It is clear that if a modest
fraction of this inflowing kinetic energy can be converted into CRs,
those CRs may become important sources of emission and even play a dynamical
role.
We also point out that $\Phi(E_k)$ is a steep function of
cluster temperature ($\propto T^\alpha, ~\alpha \sim$ 1.7 for SCDM and 1.4
for $\Lambda$CDM), spanning several orders of magnitude in 
the temperature range of the selected clusters. 
This means that if an acceleration mechanism at
accretion shocks around GCs possesses injection mechanism and
an efficiency
independent on the cluster properties (\eg mass, temperature), then 
we would expect hotter clusters to store a relatively larger amount of
nonthermal energy in the form of relativistic particles. This
should scale with $\Phi(E_k)$ and should produce consequent observational 
effects. 

\section{Discussion and Conclusions}

Accretion shock waves are the largest shocks in the
universe. We have pointed out that 
they are characterized by a complex structure yet their average
properties, such as radius and accretion velocity, show qualitative
similarities with theoretical predictions. 
The energy of accretion flows around GCs are powerful enough to account for 
the production of copious populations of CRs in the ICM. For energies
up to 2 $10^{16}$ eV CRs are efficiently trapped by the cluster magnetic field
(for $B\sim$ few $\mu$G; Berezinsky, Blasi \& Ptuskin 1997).
Different spectral components of CRs maybe responsible for the various 
nonthermal emissions observed from GCs (see \S 1). 
In addition, CR protons undergoing p-p
collisions could be responsible for a fair fraction of the diffuse gamma-ray
background (Colafrancesco \& Blasi 1998). 

In addition to accretion shocks, we have identified 
shocks propagating through the ICM and generated during the formation 
histories of clusters. 
These shocks could also be important sites of particle acceleration and must be
included when a detailed
study of the thermal and nonthermal properties of ICM is carried out.

Since there is evidence for a significant magnetic field inside GCs of the
order of a $\mu$G (Kronberg 1994) shock acceleration should be certainly 
applicable in the ICM, provided that shocks are present there.
However, the question still remains about the magnetic field strength
at the accretion shocks and therefore the viability of Fermi acceleration
mechanism there. Observations seem to impose a strict upper limit of 
$10^{-9}\mu$G on
both the regular and random component of the magnetic field outside GCs
(Vall\'ee 1990, Kronberg 1994), although Biermann \etal (1996) suggest
higher upper limits (\ltsima 1$\mu$G) at least along cosmic filaments.

We will explore these issues further in planned numerical studies 
of structure formation
in viable cosmological models taking into account explicitly CRs shock 
acceleration and transport.

\section{Acknowledgments}
We are grateful to Renyue Cen and Jeremiah P. Ostriker for providing the
$\Lambda$CDM cosmological data. FM and TWJ were supported by NSF grants
AST9616964 and INT9511654, NASA grant NAGS-5055 and by the 
Minnesota Supercomputing
Institute. HK was supported by a research development grant of Pusan National
University and DR was supported by KOSEF through grant 981-0203-011-2.
%
%
\vspace{1ex}
\begin{center}
{\Large\bf References}
\end{center}
%
Berezinsky, V. S., Blasi, P., \& Ptuskin, V. S. 1997, ApJ, 487, 529 \\
Bertschinger, E., 1985, ApJS, 58, 39 \\
Biermann, P. L., Kang, \& Ryu, D. 1996, in Proc. ICRR Symp. on Extremely High
Energy Cosmic Rays, Tokyo, ed. M. Nagano. \\
Bowyer, S., Lieu, R., \& Mittaz, J. P. D. 1997, IAU Symp. 188, The Hot
Universe, (Dordrecht: Kluwer), 52 \\
Bowyer, S., \& Bergh\"ofer, T. W. 1998, ApJ, 506, 502 \\
Cen, R. Y., \& Ostriker, J. P. 1994, ApJ, 429, 4 \\
Coalfrancesco, S. \& Blasi, P. 1998, Astroparticle Physics, 9, 227 \\
Ensslin, T. A., Lieu, R., \& Biermann, P. L. 1999, A\&A, 344, 409  \\
Kang, H, Ryu, D., Jones, T. W. 1996, ApJ, 456, 422 \\
Kang, H., Rachen, Biermann, 1997, MNRAS, 286, 257 \\
Lieu, R., Ip, W. -H., Axford, I. W., Bonamente, M. 1999, ApJ, 510, L25 \\
Miniati, F., Ryu, D., Kang, H., Jones, T. W., Cen, R. Y., \& Ostriker, J. P.
1999, in preparation \\
Ryu, D., Ostriker, J. P., Kang, H. \& Cen, R. Y. 1993, ApJ, 414, 1\\
Ryu, D., \& Kang, H. 1997, MNRAS, 284, 416\\
Ryu, D., Kang, H. \& Biermann, P. L. 1998, MNRAS, 335, 19\\
Sarazin, C. L., \& Lieu, R. 1998, ApJ, 494, L177\\
\end{document}